\documentclass[aps,prd,twocolumn,groupedaddress,nofootinbib]{revtex4-1}
\usepackage{graphicx}

\begin{document}


\title{Tuning, Ergodicity, Equilibrium and Cosmology}


\author{Andreas Albrecht}
\affiliation{University of California at Davis;
Department Of Physics\\
One Shields Avenue;
Davis, CA 95616\\
}



\begin{abstract}
I explore the possibility that the cosmos is fundamentally an
equilibrium system, and review the attractive features of such
theories. Equilibrium cosmologies are commonly thought to fail due to
the ``Boltzmann Brain'' problem. I show that it is possible to evade
the Boltzmann Brain problem if there is a suitable coarse grained
relationship between the fundamental degrees of freedom and the
cosmological observables.  I make my main points with simple toy
models, and then review the de Sitter equilibrium model as an
illustration. 
\end{abstract}


\maketitle

\section{Introduction}
\label{Sec:Intro}

The relationship between cosmology and the thermodynamic arrow of time
is a complex one.  The idea first put forward by
Penrose\cite{Penrose:1980ge} that the ``initial conditions'' of the
big bang fully account for the thermodynamic arrow of time we observe
is, after some missteps\footnote{For example Hawking 
\cite{Hawking:1985af} originally took a very different view, which was
corrected by Page~\cite{Page:1985ei} and Laflamme and Shellard~\cite{Laflamme:1987jb})}, now
widely accepted among cosmologists (see for example \cite{Guth:2011aa,Aguirre:2011aa,Albrecht:2011aa,Page:2011aa,Carroll:2011aa})
although not necessarily by Penrose\footnote{In later work (for
  example \cite{Penrose:1994de}) Penrose appears to have abandoned
  his original position in favor of a different explanation for the
  arrow of time that hypothesizes fundamentally irreversible
  physics.}. 

Guth's original paper on cosmic inflation\cite{Guth:1980zm} inspired many of
us to believe that a full understanding of the cosmos should
include an explanation of why the initial conditions for the big bang
(in this case meaning the conditions in the radiation era after
inflation) are ``typical'' or ``natural'' in some sense\footnote{An
  earlier paper by Starobinsky\cite{Starobinsky:1980te} had many of the
  feature of inflation, but had the opposite goal of motivating a
  unique and atypical solution to describe the cosmos}.  As has been
emphasized in~\cite{Page:1983uh,Albrecht:2002uz}, this expectation appears to be
in conflict with the idea that those same conditions give the universe
the low entropy start needed to explain the thermodynamic arrow of
time. Basically, low entropy means ``in an atypical part of phase
space'', so how can one expect to argue that such a state is typical?

Although a finite period of ``slow roll''
inflation\cite{Albrecht:1982wi,Linde:1981mu} has 
been found to offer an extraordinarily successful picture of the
origin of structure in the universe (see for
example\cite{Komatsu:2010fb,Ade:2013uln}), Guth's original idea of explaining that
the initial state of the big bang was natural (or not ``fine tuned'') has yet to be
realized. The focus of this paper is the pursuit of this original
goal. To achieve this goal, one has to offer a completion of the theory that describes
what happened before the finite period of inflation (or your favorite
alternative) that produced the
observed structure. ``Eternal
inflation''~\cite{Steinhardt:1982kg,Linde:1982ur,Vilenkin:1983xq,Albrecht:2013lh}
is one popular choice of a completion. It has the following
straightforward 
motivation~\cite{Albrecht:2011aa2}: If one believes the universe had the finite period of slow roll
inflation needed to produce the observed cosmic structure, in many
models a semiclassical extrapolation to earlier times would naturally take you
back into the ``self reproduction regime'' that gives eternal
inflation.
However this picture has so far been 
plagued by technical problems that prevent it from having any real
predictive power. Measure problems related to various infinities
stemming from the ``eternality'' of the model are a major problem (see
for example\cite{Guth:2007ng}). Potentially even more serious are the
challenges to even defining probability at all in such a
theory~\cite{Albrecht:2012zp}, even
if the regularization problems are solved (although see
\cite{Albrecht:2013aa,Albrecht:2014aa} for a more hopeful point of
view). Also, as discussed below, the question of
how much tuning may be involved in starting eternal inflation has not
been resolved.

This tension between the need for low entropy and the wish for
typicality has played out in a number of papers, including critiques of
inflation by Penrose~\cite{Penrose:1988mg} and later by
Coule~\cite{Coule:2002zb}. Papers examining toy models of cosmological phase space give concrete
illustrations of this point\cite{Gibbons:2006pa,Carroll:2010aj}.  Much of
this later work takes the argument further, by pointing out that 
adding an inflationary phase to the story amounts to supposing an even
lower entropy state prior to the radiation era, thus apparently
making the problem worse:  An inflationary state is clearly lower
entropy than the subsequent radiation dominated phase, given the large
entropy production during the reheating phase that connects the two.
By comparison in the standard big bang (SBB), without inflation, there
is essentially no entropy production in the radiation era all the way back to
the singularity. Thus the SBB necessarily starts out in a higher entropy state,
compared with the much lower entropy of the inflationary state.

The same issue came up in work by Dyson et al. \cite{Dyson:2002pf}
(hereafter DKS).
Those authors proposed a specific model in which the universe was
fundamentally in equilibrium, and cosmology was obtained by
fluctuations that were intrinsic to the equilibrium state. This picture gave further force to
the tension between typicality and low entropy.  For example,
in the DKS scheme the low entropy of an early inflationary state
fed in a quantitative way into an exponential disfavoring of
cosmologies with inflation vs those without.   Furthermore, the DKS
result was a concrete example of a problem known 
for over a
century\cite{Boltzmann,Eddington:1931aa,FeynmanMess,BTanthropic},
previously discussed in 
the context of modern cosmology in
\cite{Page:1983uh,Albrecht:2002uz,Coule:2002zb,Dyson:2002pf,Rees:2003aa} and later dubbed the ``Boltzmann Brain'' problem~\cite{Albrecht:2004ke}.    

The
Boltzmann Brain problem basically is the observation that an
equilibrium state strongly favors small fluctuations over large ones,
so it seems obvious that a large out-of-equilibrium universe such as
the one we observe would be highly disfavored in any model of the
cosmos based on an equilibrium state\footnote{Boltzmann Brains can be
  an issue for non-equilibrium models as well, but such models are not
  the focus of this paper.}.  This paper points out that this 
conclusion is based on simple assumptions about the relationship
between the degrees of freedom that are in equilibrium and the
cosmological observables. More subtle relationships between these two
elements of the theory could result in a successful equilibrium
cosmology that that is not undermined by the Boltzmann Brain problem. 

There are good reasons to favor an equilibrium-based cosmological
theory. Ultimately, an equilibrium theory may prove to be our only
hope to ``explain'' the state of the universe using laws of
physics. In a true equilibrium state, there is no notion of an ``initial
state''.  The system simply exists eternally, fluctuating into one
state or another with probabilities assigned by Boltzmann factors (or
whatever formula correctly expresses the statistics of that system). For
a finite system (likely to be the only situation in which such a
picture is well-defined) the system will simply cycle through
recurrences, reappearing in any given state within a finite
recurrence time (phenomena discussed in a cosmological context
in\cite{Dyson:2002pf,Albrecht:2004ke}).  One could imagine arbitrarily ``launching'' such a 
system with a particular initial state, but in the face of the
recurrences there would be no real importance to an
initial state defined in that way. 

However, a number of authors reject finite equilibrium
cosmological models as certain to fail due to the
Boltzmann Brain problem (for example see \cite{Dyson:2002pf,Carroll:2010zz}).
According to their line of thinking our best hope is truly infinite
theories (eternal inflation and variations such as \cite{Carroll:2004pn}
are examples).  Some go further and argue that only a true infinity
can be counted on to evade the 
properties of finite systems that seem to generically lead to the
Boltzmann Brain problem\cite{Guth:2011ie}, and also are hopeful
that infinite systems have a better chance of resolving the notorious
measure problems~\cite{Guth:2007ng}. I am skeptical of this line of
reasoning, particularly because infinities can be used to hide
finely tuned assumptions about initial conditions, as discussed at length in
\cite{Hernley:2013fu}.  As the system becomes larger (on the way to infinity)
the tuning of initial conditions can actually become worse.  This
effect was illustrated in \cite{Albrecht:2004ke},
where an infinite volume universe was carefully regulated as a finite
one in the limit of large volume. In that calculation the large volume
limit reduced the impact of inflation because the probability of
starting inflation (assembling all the degrees of freedom into an
inflationary state) became exponentially smaller as the volume (and
the total number of degrees of freedom) increased.  This reduction was not
sufficiently compensated by the increased total volume produced by inflation
(even though that volume was probably factored in an overgenerous
way in \cite{Albrecht:2004ke,Albrecht:2013aa,Albrecht:2014aa}). Thus
I suspect that mathematical arguments such as those presented
in~\cite{Guth:2011aa,Guth:2011ie} about infinite theories will
ultimately be seen as formal ways of re-casting high levels of tuning
in ways where it is more difficult to identify. 

I do not find my objections to infinite theories
conclusive at this point, and I certainly find such theories a worthwhile
avenue of investigation\footnote{In fact, I would not be surprised if at least
some of these ``infinite'' theories will ultimately be found to have
sufficient symmetry to be effectively finite in practice~\cite{Albrecht:2013aa,Albrecht:2014aa}.}.  The point
of this paper is not to rule out the infinite possibility, but to show
that it is not so straightforward to rule out the possibility of a
successful finite equilibrium cosmological model using arguments based
on the Boltzmann Brain problem.  

This paper is organized as follows: Section~\ref{section:tph}
introduces the ``past hypothesis'', a concept that is important to my
discussion.  Section~\ref{section:sm} makes the essential points of this paper by 
discussing two very simple ``state machine'' toy models.  The first
toy model has the essential features of a normal equilibrium system (which
{\em would} have a Boltzmann Brain problem if interpreted
cosmologically).  The second toy model has the special features needed to
avoid the Boltzmann Brain problem.  Section~\ref{section:dse}
discusses the de Sitter equilibrium cosmological model as a possible
implementation of the ideas presented in Section~\ref{section:sm}.  My
conclusions are summarized in Section~\ref{sect:conclusions}.

\section{The past hypothesis}
\label{section:tph}

The ``past hypothesis'' is entwined in an interesting way with the
main points of this paper.  This section provides a brief summary of the past
hypothesis so it can be included in subsequent
discussions. Carroll~\cite{Carroll:2010zz} 
gives a nice presentation with plenty of background\footnote{ I
  recently learned in a private communication with R. Batterman that
  the term ``past hypothesis'' might have different meanings in
  different circles.  The usage here is the same as that of Carroll.}.

The usual statistical argument for entropy increase can be sketched
like this: If a system is found in a state with sub-maximal entropy,
there will be many more states available to it with higher
entropy. Any kind of ergodic evolution that evenly explores phase space will favor
(strongly so for large systems) evolution toward a higher entropy
state. 

\begin{figure}
\includegraphics[height=2.6323in,width=3.2in]{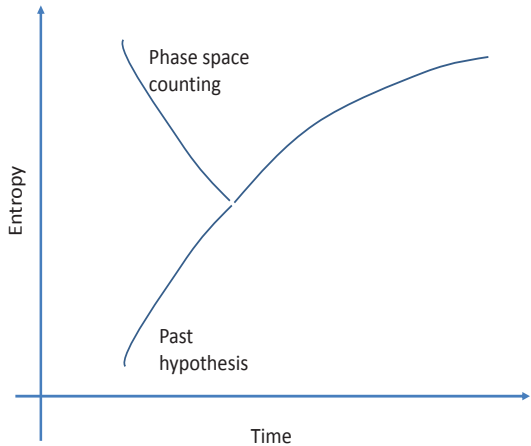}
\caption{\label{fig.DF1} Ergodic arguments allow us predict that
  entropy will increase into the future, but such state counting
  arguments would suggest that past also had higher entropy. Most
physicists find it more compelling to assume entropy decreases in the
past time direction.  This is the ``past hypothesis''.} 
\end{figure} 

One can then consider the past history of that same state.  For
usual physical laws that are time reversible, the phase space arguments
might seem to apply just as well in reverse, strongly favoring a high
entropy past as well as future. Thus one could argue that if
one finds an ergodic system in a sub-maximal entropy state at time $t_1$, it is
statistically by far most likely to be at an entropy minimum, with
entropy heading up to a higher values in both past and future time
directions. In such a picture the thermodynamic arrow of time,
pointing toward higher entropy, points {\em away} from the low entropy
state at $t_1$ in both time directions.  Figure \ref{fig.DF1}
sketches this ``double headed'' arrow of time in the upper branch.

The ``double headed'' arrow of time describes behavior we simply do not
believe to be true about our observed universe. We understand the
observed universe to have a solid thermodynamic arrow of time that
started billions of years ago and continued through to today without
reversing (at least on macroscopic scales).  This assumption is a
critical (although usually unmentioned) part of our modeling of
observable processes in the early universe (from synthesis of nuclei
to the production of the cosmic microwave background to the formation
of cosmic structure).  The success such modeling has had in matching
cosmological data is evidence for the validity of assuming the 2nd
law throughout.  This belief, that the
entropy was lower in the past, not higher as simple statistical
arguments would suggest, is the ``past
hypothesis''.  It means that
while we might appeal to some notion of ergodicity to describe the
future of a given system, we do not expect simple ergodic arguments to
describe the 
past.  Instead we believe that as we look further and
further back in time the universe occupied a smaller and smaller
region of phase space (or in other words, had a smaller and smaller
entropy), as depicted in the lower branch of Fig.~\ref{fig.DF1}. The
real universe appears to be only ``partially ergodic'': It 
is ergodic into the future but not the past. This belief has proven
itself many times over through the successes of the standard big bang
cosmology. 

The case for believing that the observed universe has a history of
increasing entropy that spans billions of years is so compelling that
it can be very difficult to imagine alternative physical situations that
do not globally respect the past hypothesis (a phenomenon
dubbed ``temporal provincialism'' in~\cite{Dyson:2002pf}).  Even papers
that discuss models in which the arrow of time is in some sense
``emergent''~\cite{Harlow:2012dd,Bousso:2011aa,Bousso:2012es} have
very strong build-in assumptions about low entropy initial conditions
(again, the past hypothesis) which are the main drivers of the arrow
of time they find to be ``emerging''~\cite{Stoltenberg:2014aa}

Since the past hypothesis is deeply tied to the 2nd law of
thermodynamics, it is also connected with the tension between 
our desire for ``typicality'' of the initial state of the cosmos and the
belief that we have experienced a multi-billion year history of
increasing entropy, requiring a finely tuned low entropy initial
state. 

\section{Ergodic state machines}
\label{section:sm}

Discussions of ergodic behavior are usually phrased in
terms of some space of microstates.  For finite systems with ergodic
dynamics, each microstate is visited for the same amount of
time, cycling through repeatedly.  The observables typically live
in a space of {\em macro}states that are related to the microstates
through coarse graining.  As discussed above, the {\em past}
behavior of the familiar microstates of  particles and fields that we usually
consider in cosmology clearly cannot be understood using ergodic state
counting arguments. In this section I use a couple of toy model
``state machines'' to illustrate come key points. 

First (Section \ref{section:sm_n}) I will illustrate a simple 
system that has fully ergodic properties.  It will be clear
that that system does not observe the past hypothesis, and were it to
be a cosmological model, it would definitely have a Boltzmann Brain
problem. 

The second state machine (Section \ref{section:sm_t}) is also ergodic,
but it has a very
different relationship between the degrees of freedom undergoing
ergodic behavior and the coarse-grained observables.  The differences
will allow the observables to respect the past hypothesis.  When
considered cosmologically, this toy model would not have a Boltzmann
Brain problem. 

I should emphasize up front that the toy models 
are nothing more than tables of numbers chosen ``by hand'' to
illustrate certain very simple points. As I will discuss below, this
utter simplicity reflects both a strength, and potentially also a 
weakness of the main points of this paper. 

\subsection{A state machine with ``normal'' fluctuations (Boltzmann Brains)}
\label{section:sm_n}

%

\begin{table*}[htbp]
  \centering

    \begin{tabular}{rrrrrrrrrrrrrrrrrrrrrrrrrrrrrrrr}

          &       &       &       &       &       &       &       &       &       &       &       &       &       &       &       &       &       &       &       &       &       &       &       & 6     &       &       &       &       &       &       &  \\

          &       &       &       &       &       &       &       &       &       &       &       &       &       &       &       &       &       &       &       &       &       &       & 5     &       & 5     &       &       &       &       &       &  \\
          &       &       &       &       &       &       &       &       &       &       &       &       &       &       &       &       &       &       &       &       &       & 4     &       &       &       & 4     &       &       &       &       &  \\
          &       &       &       &       &       &       &       & 3     &       &       &       &       &       &       &       &       &       &       &       &       & 3     &       &       &       &       &       & 3     &       &       &       &  \\
          &       &       &       &       &       &       & 2     &       & 2     &       &       &       &       &       &       &       &       &       &       & 2     &       &       &       &       &       &       &       & 2     &       &       &  \\
    Macro & 1     & 1     & 1     & 1     & 1     & 1     &       &       &       & 1     & 1     & 1     & 1     & 1     & 1     & 1     & 1     & 1     & 1     &       &       &       &       &       &       &       &       &       & 1     & 1     & 1 \\
    Micro & 1     & 2     & 3     & 4     & 5     & 6     & 7     & 8     & 9     & 10    & 11    & 12    & 13    & 14    & 15    & 16    & 17    & 18    & 19    & 20    & 21    & 22    & 23    & 24    & 25    & 26    & 27    & 28    & 29    & 30    & 31 \\
          &       &       &       &       &       &       &       &       &       &       &       &       &       &       &       &       &       &       &       &       &       &       &       &       &       &       &       &       &       &       &  \\
          &       &       &       &       &       &       &       &       &       &       &       &       &       &       &       &       &       &       &       &       &       &       &       &       &       &       &       &       &       &       &  \\
          &       &       &       &       &       &       &       &       &       &       &       &       &       &       &       &       &       &       &       &       &       &       &       &       &       &       &       &       &       &       &  \\
          &       &       &       &       &       &       &       &       &       &       &       &       &       &       &       &       &       &       &       &       &       &       &       &       &       &       &       &       &       &       &  \\
          &       &       &       &       &       &       &       &       &       &       &       &       &       &       &       &       &       &       &       &       &       &       &       &       &       &       &       &       &       &       &  \\
          &       &       &       &       &       &       &       &       &       &       &       &       &       &       &       &       &       &       &       &       &       & 3     &       &       &       &       &       &       &       &       &  \\
          &       &       &       &       &       &       &       &       &       &       &       &       &       &       &       &       &       &       &       &       & 2     &       & 2     &       &       &       &       &       &       &       &  \\
    1     & 1     & 1     & 1     & 1     & 1     & 1     & 1     & 1     & 1     & 1     & 1     & 1     & 1     & 1     & 1     & 1     & 1     & 1     & 1     & 1     &       &       &       & 1     & 1     & 1     & 1     & 1     & 1     & 1     & 1 \\
    32    & 33    & 34    & 35    & 36    & 37    & 38    & 39    & 40    & 41    & 42    & 43    & 44    & 45    & 46    & 47    & 48    & 49    & 50    & 51    & 52    & 53    & 54    & 55    & 56    & 57    & 58    & 59    & 60    & 61    & 62    & 63 \\
          &       &       &       &       &       &       &       &       &       &       &       &       &       &       &       &       &       &       &       &       &       &       &       &       &       &       &       &       &       &       &  \\
          &       &       &       &       &       &       &       &       &       &       &       &       &       &       &       &       &       &       &       &       &       &       &       &       &       &       &       &       &       &       &  \\
          &       &       &       &       &       &       &       &       &       &       &       &       &       &       &       &       &       &       &       &       &       &       &       &       &       &       &       &       &       &       &  \\
          &       &       &       &       &       &       &       &       &       &       &       &       &       &       &       &       &       &       &       &       &       &       &       &       &       &       &       &       &       &       &  \\
          &       &       &       &       &       &       &       &       &       &       &       &       &       &       &       &       &       &       &       &       &       &       &       &       &       &       &       &       &       &       &  \\
          &       &       &       &       & 3     &       &       &       &       &       &       &       &       &       &       &       &       &       &       &       &       &       &       &       &       & 3     &       &       &       &       &  \\
          &       &       &       & 2     &       & 2     &       &       &       &       &       &       &       &       &       &       &       &       &       &       &       &       &       &       & 2     &       & 2     &       &       &       &  \\
    1     & 1     & 1     & 1     &       &       &       & 1     & 1     & 1     & 1     & 1     & 1     & 1     & 1     & 1     & 1     & 1     & 1     & 1     & 1     & 1     & 1     & 1     & 1     &       &       &       & 1     & 1     & 1     & 1 \\
    64    & 65    & 66    & 67    & 68    & 69    & 70    & 71    & 72    & 73    & 74    & 75    & 76    & 77    & 78    & 79    & 80    & 81    & 82    & 83    & 84    & 85    & 86    & 87    & 88    & 89    & 90    & 91    & 92    & 93    & 94    & 95 \\
          &       &       &       &       &       &       &       &       &       &       &       &       &       &       &       &       &       &       &       &       &       &       &       &       &       &       &       &       &       &       &  \\

    \end{tabular}%
\caption{This state machine is taken to be ergodic in the microstates, stepping through them in order and spending equal times in
 each. The macrostates represent how the microstates coarse grain up
 into observables.  Macrostate $1$ represents equilibrium, and this
 model displays the expected behavior that small fluctuations are more
 frequent than large one. This table continues across three
  levels (tracking Microstates as they run from $1$-$95$). The height
  of the entries in the first row is redundant with the values, and
  is included for visual effect.    \label{tab:NoTrans}}%
\end{table*}%
Table \ref{tab:NoTrans} illustrates a very simple ``state machine''
toy model that exhibits some properties of a typical equilibrium
system. Each state is labeled by two numbers, ``Micro''
 and ``Macro'', shown one above the other.  The lower ``Micro'' label
 assigns a unique label to each state, and this row lists the complete set of
microscopic states that are accessible to the system and which are
presumed in this model to be explored ergodically in the order listed.  Since
ergodicity implies the same amount of time is spent in each state, one
can think of the ``Micro'' row as effectively a time variable. The
system is assumed to be finite, with the time evolution looping back to
microstate $1$ after the last microstate is reached. 
(One could either imagine Table \ref{tab:NoTrans} shows a complete 95
state system or just the first 95 states of a larger system with
overall similar properties.)

\begin{table}[htbp]
  \centering
    \begin{tabular}{cc}
    Macrostate & Microstates \\
    1     & 74    \\
    2     & 10    \\
    3     & 6     \\
    4     & 2     \\
    5     & 2     \\
    6     & 1     \\
    \end{tabular}%
  \caption{The frequency of appearance of different macrostates,
    listed in descending order of frequency. These counts correspond
    (through taking the logarithm) to the entropy of each
    macrostate. The highest entropy states correspond to equilibrium
    ($1$), 
    next highest participate in small fluctuations, and the lowest
    entropy states appear only in the largest fluctuations.   \label{tab:NoTransEntropy}}
\end{table}%


The ``Macro'' row represents a coarse-grained macrostate.  The index ``$1$'' 
represents the equilibrium macrostate. As one would expect of an
equilibrium state, the largest number of microstates coarse grain up to macrostate
``$1$''. One can see this by inspection of Table \ref{tab:NoTrans}.
Also, Table \ref{tab:NoTransEntropy} shows the number of microstates
that coarse grain up to each macrostate.  Macrostates with labels $>1$
represent observable fluctuations from equilibrium.  In this toy model
there are only two possible fluctuation time sequences, a small one
running up to $3$ and back down to $1$, and a large fluctuation running all
the way up to $6$ and back down to equilibrium. The large macro fluctuations
resolve into fewer microstates (again, shown in Table \ref{tab:NoTransEntropy}), and thus (thanks to ergodicity)
appear less frequently than the small fluctuations. One can think of
the large fluctuations up to $6$ as cosmological fluctuations, and the
smaller fluctuations, just up to $3$ and down as the ``Boltzmann
Brains''.  In this toy model a ``$3$'' is more likely to appear as
part of a ``Boltzmann Brain'' than as part of a cosmological state,
thus illustrating the usual problem with equilibrium-based
cosmologies (small fluctuations are more likely than large ones). Generally, the features
outlined in this paragraph correspond to features one might expect in a
realistic fluctuating equilibrium system. 

Another (closely related) way this toy model is realistic for an equilibrium system is
that it does {\em not} obey the past hypothesis. An expression of the
past hypothesis in this simple toy model might be for example, if one
finds the system in macrostate ``$3$'', it is most likely to have
been in state ``$2$'' one time step away in one time direction, but step $4$ in the
other. That is, it is part of a long-running thermodynamic arrow of
time with entropy low at one end and high on the other. In this toy model smaller fluctuations (with $2$'s on either
side of a $3$) are more favored.  Thus, while realistic for an equilibrium system, this toy model
is definitely not a good toy model for cosmology. 

There are a number of ways this toy model is {\em not} realistic
compared with everyday equilibrium physical systems.  For one,
there are only two possible fluctuations. Although the large fluctuation
does appear less frequently than the small one, no effort is made to
quantify the relative probability for the the two fluctuations in a
standard way, such as with a Boltzmann factor. It is hard
to imagine a realistic physical system that would have this rather odd
space of states.  Surely if one were found it would have to be
carefully engineered by a human to have these properties, rather than being
something easily found in nature. Still, thanks to the
realistic features mentioned in the previous paragraph, this toy model
captures enough realism for my purposes.

\subsection{A state machine with fluctuations consistent with cosmology (suppressed Boltzmann Brains)}
\label{section:sm_t}

\begin{table*}[htbp]
  \centering
    \begin{tabular}{rrrrrrrrrrrrrrrrrrrrrrrrrrrrrrrr}

          &       &       &       &       &       &       &       &       &       &       &       &       &       &       &       &       &       &       &       &       &       &       &       & 6     &       &       &       &       &       &       &  \\

          &       &       &       &       &       &       &       &       &       &       &       &       &       &       &       &       &       &       &       &       &       &       & 5     &       & 5     &       &       &       &       &       &  \\
          &       &       &       &       &       &       &       &       &       &       &       &       &       &       &       &       &       &       &       &       &       & 4     &       &       &       & 4     &       &       &       &       &  \\
          &       &       &       &       &       &       &       & 3     &       &       &       &       &       &       &       &       &       &       &       &       & 3     &       &       &       &       &       & 3     &       &       &       &  \\
          &       &       &       &       &       &       & 2     &       & 2     &       &       &       &       &       &       &       &       &       &       & 2     &       &       &       &       &       &       &       & 2     &       &       &  \\
    Macro & 1     & 1     & 1     & 1     & 1     & 1     &       &       &       & 1     & 1     & 1     & 1     & 1     & 1     & 1     & 1     & 1     & 1     &       &       &       &       &       &       &       &       &       & 1     & 1     & 1 \\
    Micro & 1     & 2     & 3     & 4     & 5     & 6     & 7     & 8     & 9     & 10    & 11    & 12    & 13    & 14    & 15    & 16    & 17    & 18    & 19    & 20    & 21    & 22    & 23    & 24    & 25    & 26    & 27    & 28    & 29    & 30    & 31 \\
    TransMicro & 1     & 2     & 3     & 4     & 5     & 6     & 7     & 8     & 9     & 10    & 11    & 12    & 13    & 14    & 15    & 16    & 17    & 18    & 19    & 20    & 21    & 22    & 23    & 24    & 25    & 26    & 27    & 28    & 29    & 30    & 31 \\
          &       &       &       &       &       &       &       &       &       &       &       &       &       &       &       &       &       &       &       &       &       &       &       &       &       &       &       &       &       &       &  \\
          &       &       &       &       &       &       &       &       &       &       &       &       &       &       &       &       &       &       &       &       &       &       &       &       &       &       &       &       &       &       &  \\
          &       &       &       &       &       &       &       &       &       &       &       &       & 6     &       &       &       &       &       &       &       &       &       &       &       &       &       &       &       &       &       &  \\
          &       &       &       &       &       &       &       &       &       &       &       & 5     &       & 5     &       &       &       &       &       &       &       &       &       &       &       &       &       &       &       &       &  \\
          &       &       &       &       &       &       &       &       &       &       & 4     &       &       &       & 4     &       &       &       &       &       &       &       &       &       &       &       &       &       &       &       &  \\
          &       &       &       &       &       &       &       &       &       & 3     &       &       &       &       &       & 3     &       &       &       &       &       &       &       &       &       &       &       &       &       &       &  \\
          &       &       &       &       &       &       &       &       & 2     &       &       &       &       &       &       &       & 2     &       &       &       &       &       &       &       &       &       &       &       &       &       &  \\
    1     & 1     & 1     & 1     & 1     & 1     & 1     & 1     & 1     &       &       &       &       &       &       &       &       &       & 1     & 0     & 0     & 0     & 0     & 0     & 0     & 0     & 0     & 0     & 0     & 0     & 0     & 0 \\
    32    & 33    & 34    & 35    & 36    & 37    & 38    & 39    & 40    & 20    & 21    & 22    & 23    & 24    & 25    & 26    & 27    & 28    & 50    & 0     & 0     & 0     & 0     & 0     & 0     & 0     & 0     & 0     & 0     & 0     & 0     & 0 \\
    32    & 33    & 34    & 35    & 36    & 37    & 38    & 39    & 40    & 41    & 42    & 43    & 44    & 45    & 46    & 47    & 48    & 49    & 50    & 51    & 52    & 53    & 54    & 55    & 56    & 57    & 58    & 59    & 60    & 61    & 62    & 63 \\
          &       &       &       &       &       &       &       &       &       &       &       &       &       &       &       &       &       &       &       &       &       &       &       &       &       &       &       &       &       &       &  \\
          &       &       &       &       &       &       &       &       &       &       &       &       &       &       &       &       &       &       &       &       &       &       &       &       &       &       &       &       &       &       &  \\
          &       &       &       &       &       &       &       &       &       &       &       &       &       &       &       &       &       &       &       &       &       &       & 6     &       &       &       &       &       &       &       &  \\
          &       &       &       &       &       &       &       &       &       &       &       &       &       &       &       &       &       &       &       &       &       & 5     &       & 5     &       &       &       &       &       &       &  \\
          &       &       &       &       &       &       &       &       &       &       &       &       &       &       &       &       &       &       &       &       & 4     &       &       &       & 4     &       &       &       &       &       &  \\
          &       &       &       &       &       &       &       &       &       &       &       &       &       &       &       &       &       &       &       & 3     &       &       &       &       &       & 3     &       &       &       &       &  \\
          &       &       &       &       &       &       &       &       &       &       &       &       &       &       &       &       &       &       & 2     &       &       &       &       &       &       &       & 2     &       &       &       &  \\
    0     & 0     & 0     & 0     & 0     & 0     & 0     & 0     & 0     & 0     & 0     & 0     & 0     & 0     & 0     & 0     & 0     & 1     & 1     &       &       &       &       &       &       &       &       &       & 1     & 1     & 1     & 1 \\
    0     & 0     & 0     & 0     & 0     & 0     & 0     & 0     & 0     & 0     & 0     & 0     & 0     & 0     & 0     & 0     & 0     & 51    & 52    & 20    & 21    & 22    & 23    & 24    & 25    & 26    & 27    & 28    & 62    & 63    & 64    & 65 \\
    64    & 65    & 66    & 67    & 68    & 69    & 70    & 71    & 72    & 73    & 74    & 75    & 76    & 77    & 78    & 79    & 80    & 81    & 82    & 83    & 84    & 85    & 86    & 87    & 88    & 89    & 90    & 91    & 92    & 93    & 94    & 95 \\
          &       &       &       &       &       &       &       &       &       &       &       &       &       &       &       &       &       &       &       &       &       &       &       &       &       &       &       &       &       &       &  \\

    \end{tabular}%
\caption{This toy model is constructed by adding a deeper layer to
  the phase space: The ``Trans-Micro'' level.  Ergodicity is
  only exhibited at this Trans-Micro level, and the Micro level is
  already a coarse graining up from Trans-Micro. Macrostates coarse
  grain up from Micro according to the same rules as for the first toy
  model. But here the coarse graining relationship between Micro and
  Trans-Micro is chosen so that large fluctuations are more frequent
  than small ones. This simple state machine illustrates the
  essential ingredients needed to build an equilibrium cosmology that
  does not suffer from the Boltzmann Brain problem.   \label{tab:Trans}}%
\end{table*}%

Table \ref{tab:Trans} shows a different toy model, with an interesting
relationship to first one.
In this model, each state has three labels. ``Macro'' and ``Micro''
are familiar from the previous model, but now there is another label,
``Trans-Micro''.  ``Trans'' designates some extension of the picture
beyond the original ``Micro'' level.  In this toy model it is the
Trans-Micro label that is assigned uniquely to each state, and
Micro now itself reflects a level of coarse-graining.  This toy
model is designed to explore the possibility that the universe is
ergodic only at a more fundamental level (the Trans-Micro level) which
describes physics beyond the ordinary microphysics of particles and fields we
usually consider when developing statistical arguments about the
universe. Specifically, I would like to explore the possibility that
ergodicity at a more fundamental level could actually be harnessed to make a
realistic cosmology (long running arrow of time and all) ``typical'',
and specifically more favored than Boltzmann Brains. 

For the toy model in Table \ref{tab:Trans}
the microstates and macrostates are the same as for first model
described in Table \ref{tab:NoTrans}. That is to say, all microstates
coarse grain up to the macro level in exactly the same way. The microstates
and macrostates are intended to represent familiar ``fundamental
particles'' and ``observables'' (respectively) in just the same way they did in the
first toy model. What is different is that here the microstates do
not evolve in an ergodic manner. 

Specifically, this state machine has been constructed so that ``large''
fluctuations (at the macro level) are more likely than small
ones. This has been accomplished in this toy model by simply
``cutting'' segments from the first toy model and ``pasting'' them on
top of the string of trans-microstates in this model.  The special
features are achieved by pasting large fluctuations more often than
small ones.  I have also allowed for the possibility that some
of the fundamental trans-microstates do not have an interpretation in
terms of microstates by
introducing the new microstate ``$0$'' to represent this possibility
(which could occur for example for highly ``stringy'' states that do
not have an interpretation as the familiar particles and fields). 

\begin{table}[htbp]
  \centering
    \begin{tabular}{cc}
    Macrostate & Trans-microstates \\
    1     & 65     \\
    2     & 8      \\
    3     & 7      \\
    4     & 6      \\
    5     & 6      \\
    6     & 3      \\
    \end{tabular}%
  \caption{The frequency of appearance of different macrostates,
    listed in descending order of the number of different trans-microstates which they coarse grain up from (and thus the frequency with which they appear in the ergodic process). These counts correspond
    (through taking the logarithm) to the entropy of each
    macrostate. The entropy of the large fluctuation states is higher than in the case shown in Table~\ref{tab:NoTransEntropy}.  \label{tab:TransEntropy}}%
\end{table}%

In this toy model macrostate $3$ is more likely to appear with a $2$
on one side and a $4$ on the other, during an extended period of
entropy increase. That is, the state $3$ is more likely to appear as
part of the evolution I have called  ``cosmological'', and less likely
to appear as part of a ``Boltzmann Brain'' fluctuation (with a $2$ on
both sides).  Thus, in 
this toy model cosmological solutions (obeying the past hypothesis)
are more likely to appear than Boltzmann Brains. This feature is
{\em due to} the ergodicity, not in conflict with it as it would be in
the first toy model (and as it would be in familiar physical
systems).  The reason for this change is that in the 2nd toy model,
the microstates no longer evolve in an ergodic way.  Only the
trans-microstates do that.  Furthermore, the features of the coarse
graining (in other words the relationships between trans-micro, micro
and macro) have been engineered to make cosmological fluctuations more
likely than small ones. 

\subsection{Further discussion of the state machines}
\label{subsection:FurtherState}

The first state machine (Table \ref{tab:NoTrans}) is meant to
represent familiar equilibrium physics.  The microstates represents the familiar
``fundamental'' particles and fields and the macrostates with index
different from unity are meant to
represent observable fluctuations away from equilibrium.  The
equilibrium behavior depicted in Table \ref{tab:NoTrans} has the
expected property that a small fluctuation is more likely than a large
one.  The segment in Table \ref{tab:NoTrans} where the microstate runs
from $24$ to $28$ represents features we believe to be true about the big bang
cosmology. The system starts in a low entropy (index $=6$) macrostate
and the entropy gradually increases.  Because of this low entropy
start (chosen to reflect the past hypothesis) one cannot
use state counting to reconstruct the past. In the cosmological
solution, the microstate corresponding to macrostate $3$ is simply
not ``typical'' according to phase space counting arguments.  It must be
a special finely tuned microstate that, when evolved backward, leads
to state with {\em even lower} entropy.  This feature of the
cosmological solution is the reason the toy model depicted in Table
\ref{tab:NoTrans} is not a good toy model for cosmology. That toy
model is in equilibrium at the micro level, so that state counting arguments do go through.
That means the cosmological behavior is disfavored over smaller
fluctuations (the Boltzmann Brains). 

The second toy model (Table \ref{tab:Trans}) is meant to illustrate a
possible extension of the model to include more fundamental degrees of
freedom. 
The extension is chosen to illustrate how one might realize a successful
equilibrium cosmology. One important feature of this extended model is
that the behavior of the macro and micro level states are
counterintuitive when interpreted using the traditional intuition we
have about familiar physical degrees of freedom (intuition which was
successfully realized in the first toy model). 

For example, consider the 
time sequences moving forward from trans-microstates with indices
$19$, $40$, and $82$. These represent the start of large fluctuations,
which in the second toy model are more likely than small ones.
In each of these cases the next step is assigned the
(coarse grained) micro index $20$. Using more traditional intuition,
this would seem to be a ``fine tuning'', since it violates state
counting arguments made in the micro space (which would say, for
example that micro index $20$ should appear no more often than any
other micro index, including micro indices $7$, $53$, $68$ and $89$,
all of which correspond to small fluctuations). The point is that this 
is {\em not} a fine tuning in the 2nd toy model, because there state counting does not work in
the micro space.  The micro space is simply a coarse-graining up from
the more fundamental trans-micro space and is not ergodic at all.  

As
I have already emphasized, by assuming the past hypothesis we have
already accepted that state-counting arguments made about the
particles and fields are not applicable to
the universe as a whole (at least when working out our past
history). The breakdown of counting arguments when applied to the
microstates of the second toy model should not be seen as a problem
with the toy model, but simply the way our (established) abandonment of counting
arguments in cosmology shows up in this toy model.  The novelty with
the second toy model is
that despite the breakdown of counting arguments at the micro level,
ergodicity and state counting can be fully recovered at the more
fundamental trans-micro level. 

The triviality of the state machine toy models deserves further
scrutiny, as it reflects both strong and weak aspects of the points I
am making here.  I have been explicit in the above discussion about
how trivial these toy models are.  They are literally just lists of
numbers I have created ``by hand''.

There is a widespread view that a profound conceptual barrier prevents successful
equilibrium cosmological models from being built, due to
issues discussed above such as the Boltzmann Brain
problem~\cite{Guth:2014aa}. The 
toy models presented here show, by trivial counterexamples, that no
such conceptual barrier exists.  Through their simplicity, the toy
models illustrate how easy it is to circumvent the conceptual issues
facing equilibrium cosmologies.  All one needs is for our elementary
particles and fields to not represent the fundamental physical phase
space. Instead, they must represent coarse-graining up from a more
fundamental phase space, and the coarse-graining relationship between the
particle-field space and the fundamental space must embody key
subtleties that let ergodicity in the fundamental space mimic tuning in
the coarse-grained one. 

The idea that fundamental physics should be expressed in a phase space
much larger than that of the everyday particles and fields is hardly
radical in contemporary physics (see for example discussions of the
string theory landscape~\cite{Kachru:2003aw}).  The more tricky issue here has to
do with the particular coarse-graining properties required to make a
successful equilibrium cosmology.  

One idea might be to imagine how one might realize the required properties in ``everyday'' physical
systems (boxes of gas with dividers, membranes, paddle wheels or
whatnot) but so far my explorations have not yielded any nice
illustration. Of course the
toy models shown in this paper are easy enough to program on a
computer, and would be just as straightforward to manufacture in a
machine shop as a mechanical device, but the 2nd toy model has
properties that seem difficult to find in more naturally occurring
systems.  Perhaps this difficulty is not a bad thing, but a
feature.  Perhaps the only possibility of realizing the special
properties of the trans-macro space is exotic degrees of freedom that are not
part of our everyday world.  Nonetheless, the fact that I have not
clearly established such an illustration means the relevance of
the points illustrated by these toy models to realistic modeling of the cosmos remains a matter
of speculation. 

\section{de Sitter equilibrium as an illustration}
\label{section:dse}
The conclusion in the previous section about the speculative nature of
equilibrium models of cosmology may seem disappointing, but it is
the state of the art for any cosmological model building.  The popular
eternal inflation model discussed in the introduction requires
speculation that the effective field theory of the inflaton is valid
over a large~\cite{Guth:2007ng} or perhaps even formally
infinite~\cite{Guth:2011ie} range of scales.  And ideas about the dynamics
of a string theory landscape, while influential, are even more
speculative.  

The de Sitter equilibrium model (dSE) is constructed by deliberately shaping
the speculation (inherent in any of the current attempts at a fundamental picture of cosmology) in
direction needed to construct a viable equilibrium theory. Whether or
not these particular speculations turn out to be correct, I find the
dSE model
interesting as an illustration of what sort of behavior might be
required of a fundamental theory to result in a successful equilibrium
cosmology.  The dSE model has been discussed at length
elsewhere~\cite{Albrecht:2004ke,Albrecht:2009vr,Albrecht:2011yg,Albrecht:2013lh}. Here I just provide a sketch, with the purpose of tying it in with the
earlier parts of this paper. 

The de Sitter equilibrium cosmology takes its inspiration from certain
``holographic'' ideas about de Sitter space, in the case where one
assumes the cosmological constant is truly fundamental and not able
to decay.  As shown in~\cite{Gibbons:1977mu} a pure de Sitter space has
maximum entropy (vs states where other objects such as black holes are
added in).  As discussed originally in~\cite{Banks:2000aa,Fischler:2000aa}, one
might be tempted to imagine the full quantum of theory de Sitter space
can be expressed in a {\em finite} Hilbert space with dimension 
\begin{equation}
N = e^{S_\Lambda}.
\end{equation}
If you never observe an entropy larger than $S_\Lambda$, why would you
need $ln(N) > S_\Lambda$?.  In such a picture the fundamental degrees
of freedom would be sufficient only to describe a single horizon
volume of space, ending at the ``thermal'' de Sitter 
horizon where fundamental quantum effects would be expected to be
important. Following similar ideas from ``black hole
complementarity''\cite{Susskind:2005js}, one might expect
transformations to the frames of 
different observers to reorganize the physical degrees of freedom in a
nonlocal manner to describe different-looking single horizon volumes,
even if there are only sufficient degrees of freedom to describe one
horizon volume at once\footnote{It is important to remember when
  considering this somewhat odd picture that most of the degrees of
  freedom, from the point of view of any one observer, are tied up in
  the ``thermal horizon''.  Some of these can swap in and out of the
  ``interior'' region of semiclassical spacetime when transforming
  from one observer to another.}.

As discussed first by DKS~\cite{Dyson:2002pf}, it seems natural to think that
such a system would exist in (or be driven to) equilibrium, undergoing
fluctuations around the 
equilibrium (de Sitter) state.  Ergodic arguments suggest that
localized fluctuations (that perturb space so that at a distance the
perturbation looks Schwartzschild with ADM mass $m_f$) would happen with
probability 
\begin{eqnarray}
P_f & = & {exp{\left(S_\Lambda - \sqrt{S_\Lambda S_m}\right)}\over
  {exp{\left(S_\Lambda\right)}}} \\
& = & exp{\left( - \sqrt{S_\Lambda S_m}\right)} \\
& = & exp{\left(-m_f/T_{GH}\right)} \label{GHfluct} 
\end{eqnarray}
where $S_m$ is the entropy of a black hole with mass
$m_f$ and 
\begin{equation}
T_{GH}=\frac{1}{2\pi}\sqrt{\frac{3}{\Lambda}}.
\end{equation}
\ Equation~\ref{GHfluct} is a result from Gibbons and
Hawking~\cite{Gibbons:1977mu} which comes from a careful analysis of the
change in area of the de Sitter horizon when the localized
perturbation is introduced. Rather remarkably, it winds up giving a
simple exponential suppression of the Boltzmann form
(Eqn.~\ref{GHfluct}), which belies the general relativistic origin of this 
result. 

In~\cite{Albrecht:2009vr} I consider fluctuations which tunnel to an
inflationary 
state with energy density $\rho_I$ and
estimate the mass of such fluctuations to be
\begin{equation}
m_s= 0.001kg \left(  { \left(  10^{16}GeV \right)^4  \over \rho_I } \right)^{1/2}.
\label{ms}
\end{equation}
I also argue that the probability of tunneling to inflation is well
approximated by Eqn.~\ref{GHfluct} with $m_f = m_s$. Thus, as
long as the universe can not produce a ``brain'' fluctuation with mass
of only a gram, it is more likely to fluctuate into a tunneling event
that creates 
an entire inflationary universe large enough to encompass all that we
see. 

So far these quantitative results are fairly standard ones, and do not
presuppose a finite system, with a finite Hilbert
space. In~\cite{Albrecht:2009vr,Albrecht:2011yg,Albrecht:2013lh} I
consider the following ``completion'' of this picture 
(speculative of course, as is the case with any other fundamental picture of
inflation)\footnote{This picture is identical to that
  in~\cite{Carroll:2004pn} except for the completion I describe
  next.  Carroll and Chen complete their picture in the form of an
  infinite theory}. 

I interpret the tunneling process as the evolution of the full quantum
state (with some suppressed tunneling probability) from one that
describes one semiclassical spacetime (the de 
Sitter space) to one that describes two semiclassical spacetimes: The
perturbed de Sitter spacetime and the ``baby
universe'' that has budded off (semiclassically speaking) and started
inflating. As the baby 
universe inflates, reheats and evolves through a standard cosmology,
the full state continues to describe both semiclassical spacetimes.
During this period, some degrees of freedom are tied up describing the
baby universe, and others are used to describe the perturbed de Sitter
spacetime. These two sets of degrees for freedom are essentially
decoupled during the baby universe phase. The
time at which the baby universe finally approaches late time de Sitter behavior (its
ultimate destiny if $\Lambda$ is fundamental) is the time at which the
degrees of freedom tied up in describing the baby universe re-couple
with those describing the perturbed de Sitter space.  Together they
form the complete set of of degrees of freedom, describing the
equilibrium de Sitter state again.  
\begin{figure}
\includegraphics[height=2.6323in,width=3.2in]{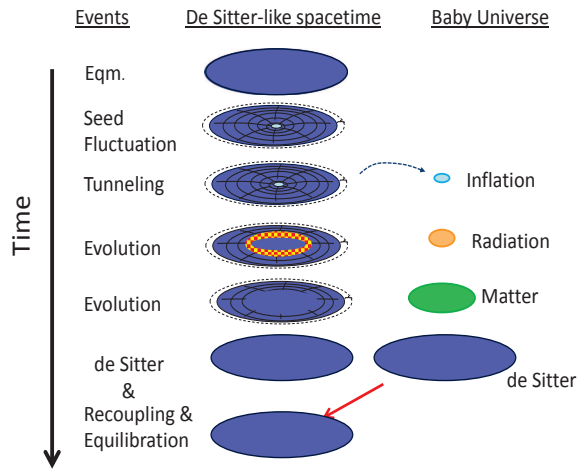}
\caption{\label{fig.DESfig} In the dSE model, the equilibrium state
  fluctuates off a baby universe which inflates, reheats and undergoes
  standard cosmological evolution eventually re-equilibrating.  The quantum state describes two
  semiclassical spacetimes between the moment of tunneling until the
  time of re-equilibration. The probability of producing this
  fluctuation is competitive with the production of an other $1$ gram
  localized fluctuation in the matter.}
\end{figure} 

During this same period, the perturbed de Sitter space will have the
following behavior:  The initial fluctuation that tunnels off the baby
universe appears, at least far away in the de Sitter space to be a
small black hole with mass $m_s$.  This small black hole will rapidly decay,
but it will take a time of order $H_{\Lambda}^{-1}$ for radiation from
the decay to reach the de Sitter horizon. During most of that time,
the decay products will appear localized, at least compared with the
size of the de Sitter horizon, and thus de Sitter horizon (and the
corresponding entropy) will remain reduced according to the Gibbons
and Hawking formula. Only after a time $~H_{\Lambda}^{-1}$
will these quantities return to their equilibrium values.
Interestingly, this is essentially the same time the baby universe will
take (according to its own cosmic time) to return to equilibrium.   The above process is sketched
pictorially in Fig.~\ref{fig.DF1}~\footnote{This picture, with it's
  repeating cosmological cycles might invite comparison with the
  ``cyclic cosmologies'' described in~\cite{Turok:2004yx}. However
  those models are envisioned as infinite systems, with infinitely
  increasing entropy~\cite{TurokPrivate}.  Thus they would seem to
  have an infinitely bad fine tuning problem for the initial state
  that starts the cycles.}. 

I now further consider the parallels between this picture and the 2nd
toy model (with the trans-macrostates) considered in 
Sect. \ref{section:sm_t}. The part of Fig.~\ref{fig.DF1} that gives the
cosmological evolution (in one of the semiclassical spacetimes)
corresponds to the large ``cosmological'' fluctuations in that toy
model.  The apparent time asymmetry of Fig.~\ref{fig.DF1} is an
artifact of the usual way of discussing quantum tunneling, and is not
an actual property of the full quantum state which includes
superpositions of the process going in both directions, as discussed
in~\cite{Albrecht:2009vr}. (The discussion in \cite{Shapere:2012nq}
also appears to be related to this point).  

Consider the following properties of the standard cosmological
evolution (with a period of early inflation): Take a snapshot
today of the microstate of all the elementary particles and fields. We
believe that microstate contains enough information that if it were
(rigorously) evolved it back in time, the high temperature early universe
state would ``de-heat'' back into an inflaton rolling back up the
hill. This is simply a feature that everyone believes to be true about
the current state of the universe, as long as one believes in an early
period of inflation. One can argue that such a state is finely tuned
in that, for example, the thermal state of the radiation era
apparently corresponds to many more microstates that will {\em not}
de-heat into cosmic inflation when time reversed.  On the other hand,
imposing the past hypothesis seems a straightforward path to admitting
such a ``tuning'' into one's theory.  So far this is just a reiteration
of points reviewed earlier in this paper. The novelty comes when one
extends this discussion more fully to the dSE model.

The time reverse of the full dSE picture (which can be seen by reading
the sketch in Fig.~\ref{fig.DESfig} from bottom to top) involves the de Sitter
horizon emitting a ``shell'' of black hole decay products with
special (highly coherent) properties  that the propagate inward and
form the hole~\footnote{
In fact, this is probably the most likely way
  to for the seed perturbation in the ``forward'' direction as well,
  so Fig.~\ref{fig.DESfig} probably should be drawn at least somewhat time symmetrically on the
  left side, with the propagating shell appearing at both ends,
  interpolating in time between seed formation/decay and
  emission/absorption of the shell at the de Sitter horizon}.  
Another special property of that event is that some
degrees of freedom decouple, producing a decoupled cosmological
spacetime that is evolving according to a time-reversed standard
inflationary cosmology.  When thought of in terms of elementary particles and
fields, this process certainly seems like a finely tuned one.  One
could compare it, for example, with the time reverse of a
(non-inflationary) standard big
bang cosmology relaxing toward de Sitter space (as was done
by DKS~\cite{Dyson:2002pf}). That would involve no splitting into two spacetimes,
and no special phase information necessary for the ``de-heating'' into
inflation discussed above.  When viewed from the point of view of
elementary particles and fields, the time reverse of Fig.~\ref{fig.DESfig}
certainly suggest that the fluctuation shown is much less likely than
one corresponding to a non-inflationary standard big bang. (This, by
the way, was the conclusion of DKS.)

This apparent fine tuning corresponds exactly to the apparent tuning in
the second toy model, where the large fluctuations occur more
frequently than the small ones.  In the case of the second toy model 
this phenomenon is actually not tuned at all, whereas it would
require tuning if
this phenomenon were observed in the first toy model.  The difference lies
in the fact that the equilibrium and ergodic behavior are taking place
in a larger set of more fundamental (trans-macro) degrees of freedom
in the second 
toy model. For that model both the macro and micro observables are
coarse grainings of the fundamental states and their behavior is
dictated only indirectly by the ergodic properties, via the special
nature of the coarse-grained relationship to the fundamental degrees
of freedom. 

For the dSE model to work, with no tunings, the system would need to
have its own equivalent to the trans-macro space that would play a similar
role. The tuning discussed above in terms of the particles and fields
could be undone in a similar way if the particles and fields where
themselves suitably coarse grained from a more fundamental set of degrees of
freedom.  The idea that the particles and fields may not be fundamental is
hardly radical these days, with the widespread hopes that ideas such as
string theory offer a better fundamental picture.  The key to the
success of the dSE model though, lies in the special nature of the
coarse graining required.  Basically the coarse graining must be
conceptually equivalent to that displayed in the second toy model for
the cosmological behavior not to be fined tuned.  At this 
point I cannot offer a specific fundamental picture which exhibits
the very special coarse graining relationship between the particles
and fields and the fundamental degrees of freedom needed to support
the dSE model.  Thus, the viability of the dSE picture remains a
matter of speculation.  One hopeful note is that this speculation
focuses on the behavior of degrees of freedom at the de Sitter
horizon, a place where as with the black hole horizon,  there seems
to be a lot we don't yet understand. 

\section{Conclusions}
\label{sect:conclusions}
The question of whether it is possible to construct a theory of
cosmology that is not finely tuned remains an open one. In this article
I have argued that cosmological theories based on equilibrium have
certain attractive features, including the fact that there are no
``initial conditions'' at all.  The probabilities of a given fluctuation
(cosmological or otherwise) are given by the laws of physics, through
the Hamiltonian as it appears in the Boltzmann factors or the
equivalent. However, equilibrium theories are notorious for the
``Boltzmann Brain'' problem, basically because they strongly favor
small fluctuations over large ones. Here I have argued on the basis of
extremely simple toy models that the Boltzmann Brain problem is not
insurmountable for equilibrium cosmological theories.  What one needs
to get around it is a special coarse graining relationship between the
familiar particles and fields and another (larger) set of more
fundamental degrees of freedom.  I have illustrated via the de Sitter
equilibrium model how such an equilibrium cosmological model might
look, but recognize that the dSE model has yet to be realized in
a fundamental theory. 

The approach I have explored here simply takes
the very special properties we know our early universe must have
and maps them onto very special properties of the coarse graining
relationship between the fundamental degrees of freedom and the
familiar particles and fields. This may seem like simply exchanging
one exotic feature for another, but the fact is that one way or
another, nature {\em has} chosen the observed universe to have some
exotic properties that look naively like tuned initial conditions. Our
job as cosmologists is to learn how nature has 
chosen to realize these features. In my view, the idea of a
fundamentally equilibrium cosmos remains an attractive contender, and
may in fact be the only alternative to simply accepting finely tuned
initial conditions as an unexplained feature of our universe.

\acknowledgements
I thank R. Bousso, D. Martin, A. Guth, D. Phillips, L. Randall,
E. Silverstein, H. Stoltenberg and L. Susskind for 
helpful discussions. I also thank the KICP and the 
University of Chicago department of Astronomy and Astrophysics for
hospitality during my sabbatical, and the Kavli Institute for
Theoretical Physics at UCSB for hospitality in June 2013.  This work
was supported in part by DOE Grants DE-FG02-91ER40674 and
DE-FG03-91ER40674 and the National Science Foundation under Grant
No. PHY11-25915. 

\bibliography{AA}

\end{document}